\documentclass[groupedaddress,amssymb,eqsecnum,epsfig]{revtex4}
\usepackage{graphicx}
\usepackage{bm}
\usepackage{dcolumn}
\usepackage{amsmath}
\def \lleq {\lower0.9ex\hbox{ $\buildrel < \over \sim$} ~}
\def \ggeq {\lower0.9ex\hbox{ $\buildrel > \over \sim$} ~}

\def \beq  {\begin{equation}}
\def \eeq  {\end{equation}}
\def \ber  {\begin{eqnarray}}
\def \eer  {\end{eqnarray}}

\begin{document}
\newcommand{\newc}{\newcommand}

\newc{\be}{\begin{equation}}
\newc{\ee}{\end{equation}}
\newc{\ba}{\begin{eqnarray}}
\newc{\ea}{\end{eqnarray}}
\newc{\bea}{\begin{eqnarray*}}
\newc{\eea}{\end{eqnarray*}}
\newc{\D}{\partial}
\newc{\ie}{{\it i.e.} }
\newc{\eg}{{\it e.g.} }
\newc{\etc}{{\it etc.} }
\newc{\etal}{{\it et al.}}
\newcommand{\nn}{\nonumber}
\newc{\ra}{\rightarrow}
\newc{\lra}{\leftrightarrow}
\newc{\lsim}{\buildrel{<}\over{\sim}}
\newc{\gsim}{\buildrel{>}\over{\sim}}
\title{Fitting Non-Minimally Coupled Scalar Models to Gold SnIa Dataset}
\author{Mingxing Luo and Qiping Su}
\affiliation{Zhejiang Institute of Modern Physics, Department of
Physics, Zhejiang University, Hangzhou, Zhejiang 310027, P R
China\\e-mail:\ luo@zimp.zju.edu.cn ,    qps@zimp.zju.edu.cn}
\date{\today}

\begin{abstract}
Non-minimally coupled theories of special potentials are analyzed numerically.
Such theories yield equations of state $\omega<-1$ and oscillations of the cosmological expansion,
which are favored by the recent analysis of observations.
Fitting these theories to the Gold SnIa dataset, we obtain results comparable with other models.
A potential of the form $V(\phi)=V_0e^{a_1\phi^2}$ yields $\chi^2_{min}=170.127$.
Similar results are obtained for potentials of the form $V(\phi)=V_0+a_1\phi^n$.
\end{abstract}

\maketitle

\section{Introduction}
It has been suggested strongly by detailed observations of distant Type Ia Supernovae (SnIa) \cite{gold} that
our universe may be in a phase of accelerating expansion.
On the other hand, measurements of cosmic microwave background (CMB) \cite{cmb} and other surveys
\cite{lss} indicate the spatial flatness of the universe.
The simplest explanation for these observations is a dominating cosmological constant, whose equation of state is
$\omega_\Lambda=p_\Lambda/\rho_\Lambda=-1$, with a complementary Cold Dark Matter (LCDM):
\be
H^2(z)=H_0^2[\Omega_\Lambda+\Omega_{m0}(1+z)^3]
\ee
where $z$ is the redshift, $\Omega_{m0}$ is the present value of matters
(including the dark matter, whose existence has well been confirmed)
and $\Omega_\Lambda$ that of the cosmological constant.
Spatial flatness demands that $\Omega_\Lambda+\Omega_{m0}=1$.

LCDM is a simple proposition
but requires extreme fine tuning of the cosmological constant $\rho_\Lambda$.
More so, its fit with the Gold dataset (157 most updated and reliable set of SnIa) is not very good.
Alternative models with evolving dark energy, such as quintessence, phantom, extended gravity,
were then put forward \cite{de}, most of which contain a rolling scalar field.
On the other hand, recent analysis of Gold dataset and other observations seem to favor
an equation of state $\omega<-1$ and oscillations of cosmological expansion \cite{state,oscillation}.

In normal theories of scalar fields, the equation of state is usually
\be
\omega_\phi=\frac{\frac{1}{2}\dot{\phi}^2-V(\phi)}{\frac{1}{2}\dot{\phi}^2+V(\phi)}
.
\ee
Obviously, it is not possible to have $\omega<-1$.
One attractive way out of this situation is to consider extended theories in which scalars are non-minimally coupled.
These extended theories appear naturally and widely in string theories and theories of extra dimensions in general.

In this paper, we consider simple theories of a non-minimally coupled scalar field $\phi$
which is oscillating around the minimal of the potential $V(\phi)$.
In this type of theories, $\omega<-1$ and oscillations of the cosmological expansion can be obtained readily,
as the oscillating scalar field is now strongly coupled to the gravity,
and it is well known that non-minimally coupled theories can violate energy conditions.

Fitting to the Gold dataset, these models yield sensible results
comparable with other models. The best fit parameters of these
models show that the oscillating field $\phi$ may account for both
the dark matter and the dark energy in the cosmological expansion.
The behavior of $\phi$ depends on the energy density, so $\phi$
could differ significantly within or without galaxies. Properly
averaged over space, $\phi$ is scale-dependent. We have
concentrated our analysis mainly on large scales, though it is
possible that the dark matter appearing on other observations is
also due to motion of the same scalar field. Interestingly, our
models which have properties of $\omega < -1$ and oscillating
cosmological expansion seem to have better fit, which can be
interpreted as circumstantial evidence for oscillations of the
cosmological expansion \cite{oscillation}. For comparison,
``$H^2(z)-z$" and ``$\omega(z)-z$" relations are plotted for our
models, LCDM model and for the best fit model OA Var.(1) in
\cite{oscillation}.

The paper is organized as follows.
In section II we present fundamentals of non-minimally coupled theories
and outline the procedure of fitting models to the Gold dataset.
We then analyze two specific models for illustrations, with results presented in detail in section III.
In section IV, we conclude with some discussions.

\section{FUNDAMENTALS OF NON-MINIMALLY COUPLED THEORY AND THE FITTING PROCEDURE}

Generically, the lagrangian density for a non-minimally coupled scalar field theory assumes the following general form
\be S=\int
d^4x\sqrt{-g}\left[\frac{1}{2}F(\phi)R - \frac{1}{2}W(\phi)
\partial^{\mu}\phi\partial_{\mu}\phi - U(\phi) + L_N\right]
\label{la}
\ee
where $L_N$ is the total lagrangian density of fields and matters in the universe other than the field $\phi$.
In this paper, the theory is parameterized such that $W(\phi)=1$ and $F(\phi)$ of the simple form
\be
F(\phi)=1-\xi\phi^2
\ee
here we have set $8\pi G=1$.
Assuming a flat Friedmann-Robertson-Walker metric, i.e., we assume a flat prior in all calculations of this paper,
the Einstein equations are \cite{scalar1,scalar2}:
\ba
 H^2 &=& \frac{1}{3F}\left(\rho_N
+{1\over 2} \dot\phi^2 + U - 3 H \cdot \dot F\right) \label{b}
\\
\dot H  &=& -\frac{1}{2F}\left[(\rho_N+p_N) + \dot \phi^2 +\ddot F
- H\cdot \dot F\right] \label{c} \\\nonumber
\ea
where $\rho_N$ and $p_N$ come from the $L_N$, representing, respectively,
the energy density and the pressure of matters and fields in the universe other than $\phi$.
In what follows, we will approximate these matters and fields as a perfect fluid with $p_N=0$.

For simplification, one makes the following redefinitions
\ba
 Q=H^2/H_0^2, \ \ \ V(\phi)&=&U(\phi)/H_0^2
\ea
where $H_0$ is the present value of expansion rate $H$.
From Eqs (\ref{b}) and (\ref{c}), one obtains \cite{new,scalar}
\begin{widetext}
\ba
F'' + \left[\frac{Q'}{2Q}-\frac{4}{1+z}\right]~F' &+&
\left[\frac{6}{(1+z)^2} - \frac{2}{(1+z)}\frac{Q'}{2Q}\right]~F - \frac{2 V}{(1+z)^2 Q} - 3 \frac{1+z}{Q}
\Omega_{N0}=0\  \label{d} \\
Q(z)&=&\frac{V/3+(1+z)^3\Omega_{N0}}{F-F'(1+z)-\frac{\phi'^2}{6}(1+z)^2}
\label{e} \ea
\end{widetext}
where $\Omega_{N0}={\rho_{N0}}/{3H_0^2}$.
In these two equations, we have replaced derivatives with respect to time by those with respect to the redshift $z$,
which are in turn denoted by primes.
Substitution of Eq (\ref{e}) into Eq (\ref{d}) yields a second-order differential equation which includes only field $\phi$.

To be consistent with measurements in the solar system \cite{solar},
stringent constraints have to be put on these theories.
Defining a post-Newtonian parameter\cite{gr}
\be
\gamma-1=-\frac{(dF/d\phi)^2}{F+(dF/d\phi)^2}\ \  ,
\ee
solar system tests give the following (present) upper limit \cite{solar}
\be
|\gamma -1 | < 4 \times 10^{-4}, \ \ \ \
\ee
which is equivalent to
\be
\frac{1}{F}\left( {dF \over d\phi } \right)^2_0<4\times10^{-4} \ \Rightarrow \ |\xi\phi(0)|<10^{-2}
\ee
To generate suitable oscillations in the cosmological expansion,
$\xi$ should be of $O(1)$ (see the next section).
The constraint is now converted into a constraint on $\phi(0)$.
Numerically, $\phi(0)<10^{-3}$ is strict enough.

Tests in the solar system constrain only the behaviors of $\phi$ in our galaxy.
According to its equation of motion
\be
\partial^2\phi-V_{,\phi}+\frac{1}{2}F_{,\phi}R=0
\ee (where $R$ is the Ricci scalar) the evolution of $\phi$
depends on the energy density, for $R$ is proportional to the
energy density. Usually, the potential $V(\phi)$ is tiny, such as
the ones appearing in the next section. Within galaxies, where the
density of matters is much larger compared with the average
density of the universe, the potential term in the above equation
can be neglected. Outside of galaxies or averaged over large
scales, $V(\phi)$ plays the main role and the frequency of
oscillations of $\phi$ would be much smaller. So the magnitude and
the evolution of $\phi$ are scale-dependent and sensitive to
positions, which produce varied effects at small and large scales.
As we are interested in cosmological expansion at large scales,
$\phi(0)$ is much less constrained.

For simplicity, we will take $\phi(0)=0$ as one initial condition.
Other proper choices of $\phi(0)$ will yield qualitatively similar results.
On the other hand, the effective Newton constant
\be
G_{\rm eff}=\frac{1}{F}\frac{2F+4(dF/d\phi)^2}{2F+3(dF/d\phi)^2}
\ee
changes in time, which is also constrained by present observations
\be
\left| {\dot G_{\rm eff} \over G_{\rm eff} }\right|_{z=0} < 6 \times 10^{-12} {\rm yr}^{-1}
\ee
By choosing the particular initial condition $\phi(0)=0$, one automatically has
\be
\left| {\dot G_{\rm eff} \over G_{\rm eff} }\right|_{z=0} =0
\ee
The second initial condition is obtained by evaluating Eq (\ref{e}) at $z=0$:
 \be
\phi'^2(0)=6-2V(0)-6\Omega_{N0} \label{g}
\ee
From Eqs (\ref{d}), (\ref{e}) and these initial conditions,
one obtains the current equation of state
\be
\omega(0)={1-4 \xi\over 6} \phi'^2(0) - {V(0) \over 3}
\ee
If $\xi$ is large enough, $\omega(0)$ can be less than $-1$ easily.

Given these two initial conditions, the second-order differential equation of $\phi$ can be solved,
at least numerically, once the exact form of $V(\phi)$ and the value of $\Omega_{N0}$ are also known.
Substituting the solution $\phi(z)$ back into Eq (\ref{e}),
we get $H(z)$ for arbitrary redshift $z$.
Thus, we solve the problem completely. It is then straightforward to compare the theory with observations.

To be definitive, one calculates the goodness of fit of the theory with the observed Gold dataset:
\be
\chi^2(\bar{M}, H)=\sum_{k=1}^{157}\frac{\left(m^{ob}(z_k)-m^{th}(z_k, H, \bar{M})\right)^2}{\sigma_{m^{ob}(z_k)}^2}
\label{h}
\ee
where
\be
m^{th}(z, H, \bar{M})=\bar{M}+5\log_{10}\left((1+z)\int_0^z dz' \frac{H_0}{H(z')}\right)
\ee
is the apparent magnitude and
\be
\bar{M}=M+5\log_{10}(\frac{c H_0^{-1}}{Mpc})+25
\ee
is the magnitude zero point offset.

\begin{table*}[t!]
\begin{minipage}{0.88\textwidth}
\caption{Best fit parameters and error bars for two types of potentials in non-minimally coupled (NMC) theories,
compared with the LCDM model and the OA Var. model in [6].
The first two rows are for the first form of potential.
Shown in the first row are the results with $\Omega_{N0}$ as a freely fitting parameter,
while shown in the second row are those by setting $\Omega_{N0}=0.045$.
The next three rows are for the second form of potential, with power index $n=2,\ 4,\ 6$, respectively.
When estimating the error for each parameter, all other parameters are fixed at their best fit values,
except that $\Omega_{N0}$ is fixed at $0.045$ in the second and the fifth rows.
}
\end{minipage}\\
\begin{tabular}{ccccccc}
\hline
\hline\\
\bf{Model} &   $\bf{V(\phi)}$ \bf{or} $\bf{ H^2(z)}$
&$V_0$&$a_1$&$\bf{\xi}$& $\bf{\Omega_{N0}}$  & $\bf{\chi^2_{min}}$
\\
\\\hline \vspace{-5pt}\\
NMC E(0) & $V(\phi)=V_0e^{a_1\phi^2}$ &$2.18\pm0.08$&$25.9^{+3.9}_{-2.6}$& $1.94\pm0.67$& $0.066^{+0.16}_{-0.021}$&170.127 \\
\\
NMC E(1) & $V(\phi)=V_0e^{a_1\phi^2}$ &$2.19\pm0.08$&$25.0^{+4.1}_{-2.5}$& $1.79^{+0.61}_{-0.63}$& $0.045$ & 170.135 \\
\\
NMC M2(2) &$V(\phi)=V_0+a_1\phi^2$&$2.05\pm0.015$&$-34.7^{+5.9}_{-3.7}$&$16.0^{+2.0}_{-1.2}$&$0.300^{+0.006}_{-0.004}$& 171.870 \\
\\
NMC M4(3) &$V(\phi)=V_0+a_1\phi^4$&$2.30\pm0.08$&$144.6^{+69.4}_{-44.6}$&$0.50\pm0.18$&$0.078^{+0.050}_{-0.033}$&172.494\\
\\
NMC M6(4) &$V(\phi)=V_0+a_1\phi^6$&$2.67^{+0.03}_{-0.04}$&$18.6^{+22.4}_{-11.2}$&$0\pm0.04$&$0.045$&173.963\\
\\
OA Var.(5)& \ \ \ $H^2(z)=H_0^2 [0.3(1+z)^3+0.7+0.13$ \ \ \ &&&&&\\
&$ \ \ \ (1+z)^3[\cos{(6.83z+4.57\pi)}-\cos{(4.57\pi)}] ] $ \ \ \ &&&&&171.733\\
\\
LCDM(6)&$H^2(z)=H_0^2[0.31(1+z)^3+0.69]$&&&&&177.072\\
\\
\hline \hline
\end{tabular}
\end{table*}
To find the best fit parameters of the model, one minimizes $\chi^2$.
Since $\chi^2$ is a second-order polynomial of $\bar{M}$,
the minimization procedure with respect to this parameter is straightforward and simple.
The final result of $\chi^2_{min}$ can be expressed in terms of quantities independent of $\bar{M}$
\cite{oscillation}:
\be
\chi^2(H)=A-B^2/C
\label{chimini}
\ee
where
\ba
A &=& \sum_{k=1}^{157}\frac{\left(m^{ob}(z_k)-m^{th}(z_k, H, \bar{M}=0)\right)^2}{\sigma_{m^{ob}(z_k)}^2}\nonumber \\
B &=& \sum_{k=1}^{157}\frac{\left(m^{ob}(z_k)-m^{th}(z_k, H,\bar{M}=0)\right)}{\sigma_{m^{ob}(z_k)}^2}  \\
C &=& \sum_{k=1}^{157}\frac{1}{\sigma_{m^{ob}(z_k)}^2} \nonumber
\ea
Eq (\ref{chimini}) is the starting point of our numerical analysis.
By minimizing this quantity, one obtains the best fit values of relevant parameters.
As we shall see in the next section,
both $\omega(0)<-1$ and oscillations of cosmological expansion are favored by experiments.

\section{NUMERICAL RESULTS}

Taking two special potential forms for illustration,
we now calculate the fit of non-minimally coupled theory with the Gold dataset
and find the best fit parameters in the process.
In order to obtain acceleration and oscillations of the cosmological expansion simultaneously,
we consider potentials that have non-zero minima $V_0$ at the point $\phi=0$.
Though we have only analyzed two forms of potentials in this paper,
our analysis can easily be extended to other potentials.
Other potentials of this type could yield comparable fitting results,
as in general one can easily obtain appropriate oscillations and acceleration of the cosmological expansion
from this type of models.
All results are shown in Table I.
The errors are obtained with an increase of $\chi^2_{min}$ by 1,
while we have assumed that the value of $\Omega_{N0}$ is at lest $0.045$,
which is the density of all know matters including baryons and CMBR.
When estimating the error for each parameter, all other parameters are fixed at their best fit values.
For comparison and concreteness, we will plot the evolving of $\phi(z)$,
$H^2(z)$ and $\omega(z)$ for all models in  FIG. \ref{phi}, FIG. \ref{h2} and FIG. \ref{omega1}, respectively.

\subsection{$V(\phi)=V_0e^{a_1 \phi^2}$}

This potential appears naturally in supergravity theories (SUGRA) \cite{sugr}. This form
was also reconstructed from SnIa Gold dataset in \cite{new},
where it is only a good fit of the polynomial parametrization model in reference \cite{oscillation}
and $\phi$ is not oscillating. In total, we have four free parameters: $\xi,\Omega_{N0},V_0$ and $a_1$.
Their best fit values are shown in first line of Table I (NMC E(0)).

\begin{figure}[h]
\includegraphics[width=8cm,height=5cm]{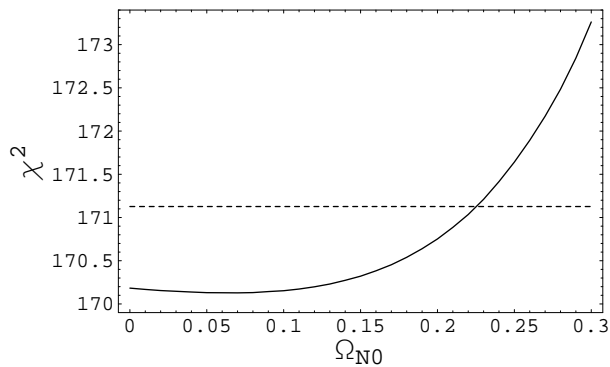}
\caption{The dependence of $\chi^2$ on $\Omega_{N0}$ with other parameters freely,
and the dash line is for $\chi^2_{min}+1$.}
\label{f1}
\end{figure}

FIG. \ref{f1} shows the dependence of $\chi^2_{min}$ on $\Omega_{N0}$, when other parameters are allowed to vary freely.
The best fit of $\Omega_{N0}$ is about $0.066$.
However,  Fig. \ref{f1} indicates that one gets almost no change in $\chi^2$ if $\Omega_{N0}$ is set to be $0.045$.
The latter is the value of all known matters including baryons and CMBR \cite{baryon}.
That is to say, the existence of a single non-minimally coupled scalar field $\phi$ may play both the roles
which dark matter and dark energy would play in LCDM on the cosmological expansion.
Of course we can split this effect into two parts:
dark matter (which also generates oscillations of the cosmological expansion) and dark energy.
Clearly, this splitting is rather artificial, as there are no obvious distinctions between these two parts in this model.
They can just be regarded as a whole as the expansion of the universe is concerned.
Now that there is no isolated dark energy, we plot in FIG. \ref{omega1} the total equation of state of the universe.
Dark matters from other observations can also come from this field \cite{galaxy},
though the behavior of the field should be scale-dependent \cite{dd}.
In this paper we only consider the effect of the field $\phi$ on cosmological expansion at large scales.

Clearly, the bigger $\Omega_{N0}$ is, the worse is the fit to Gold dataset.
This may be seen as follows.
To obtain enough amplitude of oscillations of field $\phi$,
i.e., to obtain enough oscillations of cosmological expansion to fit observations,
$\phi'^2(0)$ should be large enough.
On the other hand, to have a dominating dark energy, $V_0$ should be positive and large enough.
According to Eq (\ref{g}),  $\Omega_{N0}$ cannot be too larger.
In the rest of this subsection, we will fix $\Omega_{N0}=0.045$ to analyze other parameters.
A re-analysis of this model with this value of $\Omega_{N0}$ is given in the second row of Table I.

\begin{figure}[h]
\includegraphics[width=8cm,height=5cm]{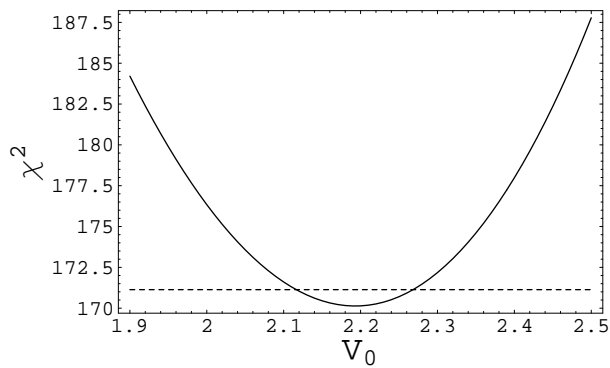}
\caption{$\chi^2$ v.s. $V_0$ with other parameters fixed as: $\Omega_{N0}=0.045, a_1=25.0, \xi=1.79$,
and the dash line is for $\chi^2_{min}+1$. }
\label{f2}
\end{figure}

The parameter $V_0$ here plays somewhat the role of cosmological constant,
with the effective $\Omega_{\Lambda}=V_0/3$.
Our best value $V_0 \sim 2.19$ is consistent with the best fit value $\Omega_{\Lambda}=0.69$ in LCDM model.
If $V_0$ departs too much from the best fit value, we will have a comparatively large $\chi^2$,
which is confirmed from FIG. \ref{f2}.
In FIG. \ref{f2} we have plotted the $\chi^2-V_0$ relation by fixing the other parameters as:
$\Omega_{N0}=0.045$, $\xi=1.79$ and $a_1=25.0$.

\begin{figure}[h!]
\includegraphics[width=8cm,height=5cm]{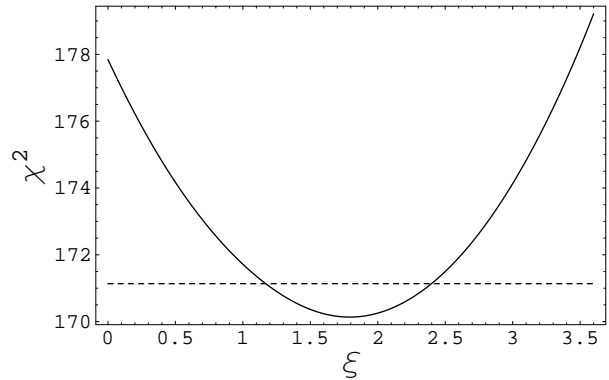}
\caption{$\chi^2$ v.s. $\xi$ with $\Omega_{N0}=0.045, V_0=2.19, a_1=25.0$.
The dash line is for $\chi^2_{min}+1$. }
\label{f4}
\end{figure}

Now we turn to the parameter $\xi$, which represents the coupling strength between gravity and $\phi$.
As $\phi$ oscillates around the minimum of potential $V(\phi)$,
$\xi$ affects the amplitude of oscillations of cosmological expansion.
If $\xi$ is very small, there will be almost no oscillation appearing in the expansion rate $H^2$ and
it is difficult to cross the $\omega=-1$ line.
Its behaves nearly the same as that of the minimally coupled theory \cite{scalar1}.
If $\xi$ is very larger, the motion of the universe will be drastically changed.
So the value of $\xi$ should be moderate.
In fact, the best fit value of $\xi$ is of the order $O(1)$ (see Fig. (\ref{f4})).

\begin{figure}[h!]
\includegraphics[width=8cm,height=5cm]{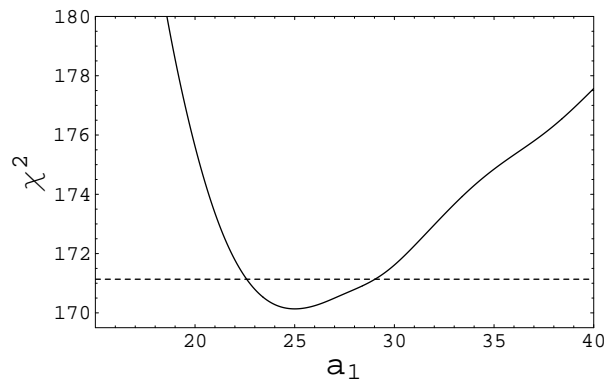}
\caption{$\chi^2$ v.s. $a_1$ with $\Omega_{N0}=0.045, V_0=2.19, \xi=1.79$.
The dash line is for $\chi^2_{min}+1$.}
\label{f3}
\end{figure}

Last but not the least, the parameter $a_1$ can affect frequency of oscillations,
its effects on $\chi^2$ are shown in FIG. \ref{f3}.
As mentioned above, we have plotted $\phi(z)$, $H^2(z)$, and $\omega(z)$ in terms of $z$ for all models,
in Figs. \ref{phi}, \ref{h2} and \ref{omega1}.
One clearly sees oscillations of $\phi$ and of the cosmological expansion over time in this type of model.
The universe may well be just in a period when the $\omega$ is in the valley
and later on will transit to a period of deceleration slowly.

\subsection{$V(\phi)=V_0+a_1\phi^{n}$}

\begin{figure}[b!]
\includegraphics[width=7.4cm,height=16cm,angle=-90]{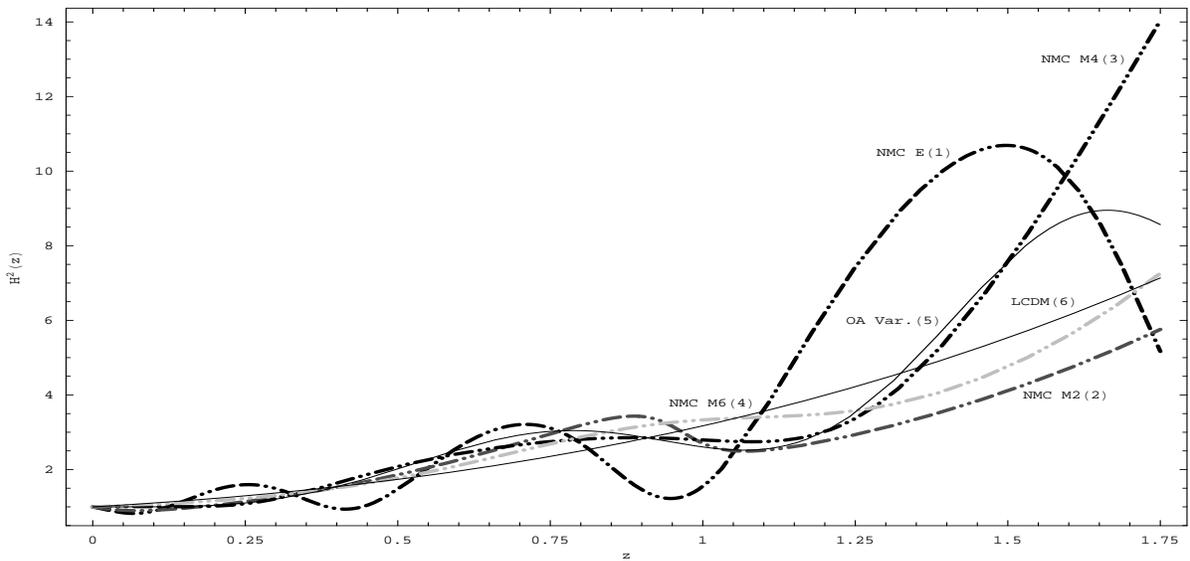}
\begin{minipage}{\textwidth}
\caption{Expansion rate $H^2(z)$ v.s. $z$ in six representative models.
Here $H(0)$ has been normalized to be unity.} \label{h2}
\end{minipage}
\end{figure}

\begin{figure}[t!]
\includegraphics[width=7.4cm,height=16cm,angle=-90]{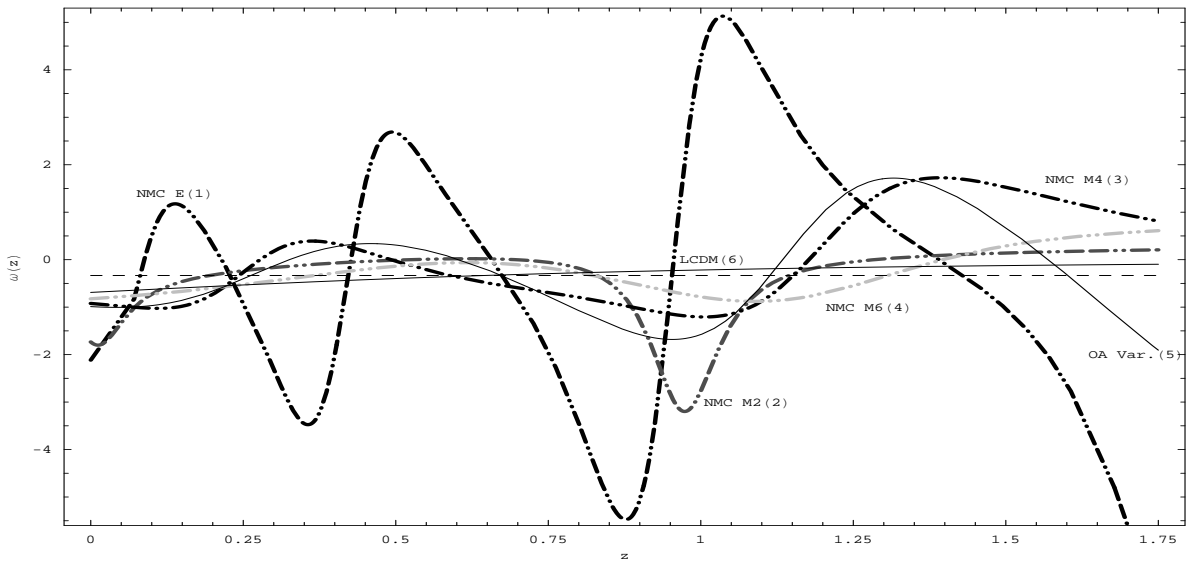}
\begin{minipage}{\textwidth}
\caption{$\omega(z)$ v.s. $z$ for the same six models.
The dashed line represents the position $\omega=-\frac{1}{3}$.
Note that it is the whole equation of state of the universe, not that just of the dark energy.}
\label{omega1}
\end{minipage}
\end{figure}

This form of potentials can be regarded as an extension of a constant dark energy,
i.e., a cosmological constant $V_0$ is added with a field $\phi$ of power potential.
The non-minimally coupled scalar field mimics the effects of dark matter
and produces oscillations in the expansion rate $H^2$.
We have considered cases of $n=2,\ 4,\ 6$, with their best fit values of parameters shown in Table I.
As shown in FIG. \ref{h2} and FIG. \ref{omega1},
the expansion patterns of these models are similar to those of LCDM
and oscillate around the line of LCDM.

\begin{figure}[b!]
\includegraphics[width=7.4cm,height=16cm,angle=-90]{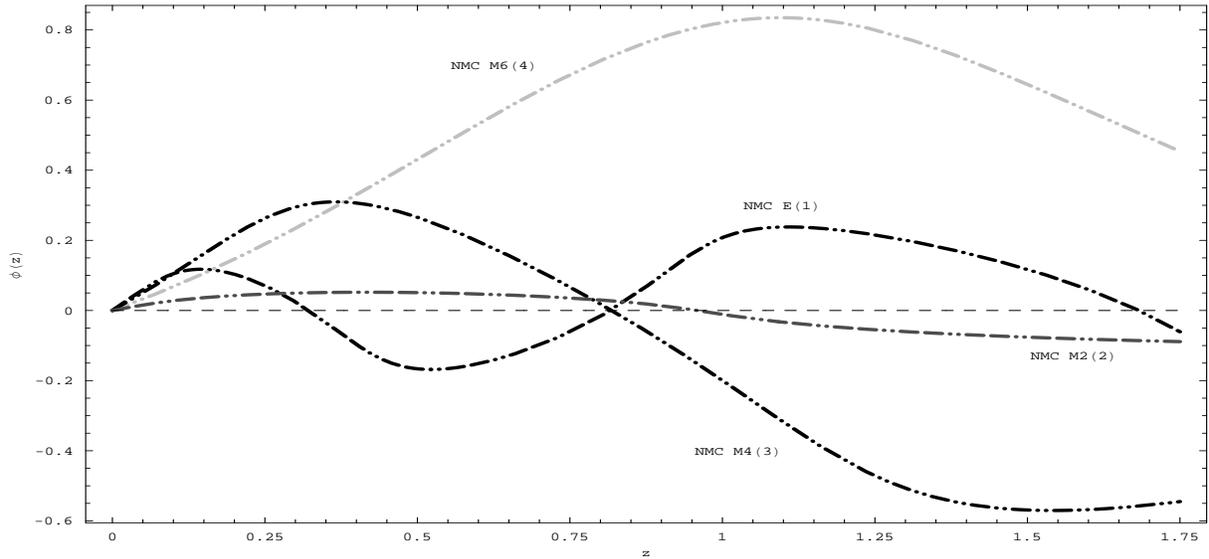}
\begin{minipage}{\textwidth}
\caption{Field $\phi(z)$ v.s $z$ in NMC models.}
\label{phi}
\end{minipage}
\end{figure}

For $n=2$, the best fit value $a_1<0$ and the potential $a_1\phi^2$ is negative,
but coupling of the scalar field with gravity can drive up the potential.
The field $\phi$ is still oscillating around $\phi=0$
if all parameters are within certain range around the best fit values.
In this model the  constraints on parameters are quite strict, as shown in Table I.
In Figs. \ref{phi}, \ref{h2} and \ref{omega1},
we see that oscillations of the cosmological expansion only appear near the $\phi=0$ points,
while in other places there is essentially no oscillation.

For $n=4$ and more so for $n=6$, the cosmological constant $V_0$ is quite large.
Then there is not much dark matter, which seems to be in conflict with other observations of dark matter.
It is possible that we can have enough dark matters at relatively small scales,
as the effect of $\phi$ is scale-dependent.
For $n=6$, the best fit value of $\xi$ is $0$,
but there are still oscillations in the expansion rate $H^2$, with a much smaller amplitude.
Its best fit value of $\Omega_{N0}$ is negative, much smaller than $0.045$, the value of known matters.
We set the value $\Omega_{N0}=0.045$ in Table I by hand.
$\,$

\section{Conclusion}

We have numerically analyzed the expansion of the universe
in non-minimally coupled scalar field theories with two special forms of potentials:
$V_0e^{a_1\phi^2}$ and $V_0+a_1\phi^n$.
These types of potentials have a no-zero minimum at $\phi=0$,
which generates the necessary acceleration of the cosmological expansion.
On the other hand, the oscillating $\phi$ produces oscillations of the expansion
and yield an equation of state $\omega<-1$, as favored by the fit with 157 observed SnIa Gold dataset.
We have further compared our models with the Gold dataset to find the best fit parameters.
The best fit potential is of the form $V_0 e^{a_1 \phi^2}$ with $\chi^2_{min}=170.127$.
The best value of all matter density in universe except $\phi$ is found to be $\Omega_{N0}=0.066$,
which is very close to all energy density of the known matters including baryons and CMBR.
So the non-minimally coupled scalar field $\phi$ may provide the main source of dark matter as well as dark energy.
Of course further investigation will be needed to make these assertions more definitive.

{\bf Acknowledgments:} This work is supported in part by the National Science Foundation of China (10425525).

\end{document}